\documentclass[aps,prb,tightenlines,twocolumn, amsmath,amssymb,superscriptaddress]{revtex4-1}
\usepackage{graphicx}
\usepackage{dcolumn}
\begin{document}
\hyphenation{spectro-meter}
\hyphenation{micro-scopy}
\hyphenation{image}
\hyphenation{Na-tu-rally}
\title{Optical spectroscopy of interlayer coupling in artificially stacked MoS$_2$ layers}
\author{G.\ Plechinger}
\affiliation{Institut f\"ur Experimentelle und Angewandte Physik,
Universit\"at Regensburg, D-93040 Regensburg, Germany}
\author{F.\ Mooshammer}
\affiliation{Institut f\"ur Experimentelle und Angewandte Physik,
Universit\"at Regensburg, D-93040 Regensburg, Germany}
\author{A. Castellanos-Gomez}
\affiliation{Instituto Madrile\~{n}o de Estudios Avanzados en Nanociencia (IMDEA-Nanociencia), 28049 Madrid, Spain}
\author{G. A. Steele}
\affiliation{Kavli Institute of Nanoscience, Delft University of Technology, 2628 CJ Delft,
The Netherlands}
\author{C.\ Sch\"uller}
\affiliation{Institut f\"ur Experimentelle und Angewandte Physik,
Universit\"at Regensburg, D-93040 Regensburg, Germany}
\author{T.\ Korn}
\email{tobias.korn@physik.uni-regensburg.de}
\affiliation{Institut
f\"ur Experimentelle und Angewandte Physik, Universit\"at
Regensburg, D-93040 Regensburg, Germany}
\begin{abstract}
We perform an optical spectroscopy study to investigate the properties of different artificial MoS$_2$ bi- and trilayer stacks created from individual monolayers by a deterministic transfer process. These twisted bi- and trilayers differ from the common 2H stacking in mineral MoS$_2$ in the relative stacking angle of adjacent layers and the interlayer distance.  The combination of Raman spectroscopy, second-harmonic-generation microscopy and photoluminescence measurements allows us to determine the degree of interlayer coupling in our samples. We find that even for electronically decoupled artificial structures, which show the same valley polarization degree as the constituent MoS$_2$ monolayers at low temperatures, there is a resonant energy transfer between individual layers which acts as an effective luminescence quenching mechanism.
\end{abstract}
\maketitle
\section{Introduction}
In addition to graphene, a large family of two-dimensional crystal structures~\cite{Novoselov26072005,Wang_2D_review, Heine_Atlas} has attracted scientific interest in recent years. One of the most intriguing aspects of two-dimensional crystals is the possibility to  stack individual layers on top of each other to create artificial structures. In these structures, different material classes, such as insulators and semiconductors, can  be combined to form heterostructures~\cite{Novoselov_2D_hetero,Geim_hetero}. Interlayer interaction in these heterostructures may be used to break the crystal symmetry and create a superlattice potential modulation~\cite{Graphene_Butterfly}. Depending on the alignment of energy levels in adjacent layers, charge separation can occur~\cite{Kosmider13}, and interlayer excitons may form~\cite{Fang29042014}.  However, artificial stacks built from a single material  may also differ substantially from the bulk properties: the stacking process yields additional degrees of freedom, such as the relative crystallographic orientation of adjacent layers and the interlayer distance. These degrees of freedom may be exploited to systematically influence the effective band structure, especially in materials whose electronic structure  changes strongly with the number of layers. A material system that is particularly suited for this is MoS$_2$, the most prominent member of a family of semiconducting dichalcogenides that also includes WS$_2$, WSe$_2$ and MoSe$_2$~\cite{Tonndorf:13}. While bulk MoS$_2$ is an indirect-gap semiconductor, its band structure changes with the number of layers, and single-layer MoS$_2$ has a direct gap due to confinement effects~\cite{Eriksson09,Heinz_PRL10,Splen_Nano10,Heine_PRB11}. Due to the two-dimensional confinement and reduced dielectric screening in single layers, excitons with large binding energies dominate the optical properties of the semiconducting dichalcogenides, even at room temperature~\cite{Heinz_Trions,Chernikov_Rydberg_PRL14,He_PRL14}. An important property of monolayer (ML) dichalcogenides is the coupling of spin and valley degrees of freedom~\cite{Xiao12}, which is directly related to the inversion asymmetry of the ML crystal structure. Valley physics in ML dichalcogenides are directly accessible by interband transitions, as  the optical selection rules for MLs allow for valley- and spin-selective excitation of electron-hole pairs.
An optically generated valley polarization manifests itself as a partial circular polarization of the photoluminescence (PL)~\cite{Yao12,Heinz12,Zeng_WS2_12, Wu_vaporTransport, Lagarde14,Heinz_Review_Valley}. By contrast, for bilayers that are stacked in the 2H stacking order prevalent in mineral MoS$_2$, which contains an inversion center,  spin-valley coupling is absent.
Recently, artificially stacked MoS$_2$ bilayers have been studied by several groups~\cite{ACG14,Hone_Twist_NanoLett14,Wang_Twist_NatComm14,Dresselhaus_twisted_NL, Jiang_stacked_Valley_NatNano14}, and a significant drop of the PL intensity with respect to isolated monolayers was observed in these structures. Additionally, PL emission at energies below the A exciton transition in monolayers was reported, indicating a modification of the electronic band structure towards an indirect band gap in the artificial bilayers caused by interlayer coupling.

Here, we use optical spectroscopy to investigate the properties of different artificial MoS$_2$ bi- and trilayers created by a deterministic transfer process. The combination of Raman spectroscopy, second-harmonic-generation microscopy and photoluminescence measurements allows us to determine the degree of interlayer coupling in our samples. We find that we are able to create decoupled artificial bilayers and trilayers, which demonstrate the same valley polarization degree as the MoS$_2$ monolayers at low temperatures, indicating that the coupling is weak and that the inversion asymmetry of the monolayer is preserved. Although the coupling in our artificial structures is weak enough to preserve the valley polarizability and direct-gap band structure of isolated monolayers,  we are able to identify a resonant energy transfer between layers, which acts as a quenching mechanism for the photoluminescence. Its efficiency strongly depends on the interlayer distance and on the photocarrier dynamics, which are modified by changing the sample temperature.
\section{Methods}
\subsection{Sample preparation}
We use two different ways of creating  artificial MoS$_2$ stacks. Both techniques are based on a recently developed deterministic transfer process~\cite{Gomez_Transfer}: first, mono- and few-layer MoS$_2$ flakes are created from  bulk MoS$_2$ mineral crystals by mechanical exfoliation using adhesive tape.   The flakes are then deposited on viscoelastic, transparent polydimethylsiloxane (PDMS) substrates. Subsequently, the flakes are transferred from the PDMS onto a silicon substrate. Here, we use a silicon wafer with a 285~nm thick SiO$_2$ oxide layer and predefined metal markers. During the exfoliation and transfer to the PDMS, folds develop in some of the flakes, leading to the random formation of twisted bi- and trilayer areas. These folded areas are typically rather small (a few square $\mu$m), but sufficiently large for optical spectroscopy measurements. Due to the fact that the folding process occurs during the exfoliation, the flake surfaces within the fold are only directly exposed to ambient conditions for a short period of time.

Alternatively, we can prepare artificial stacks  of MoS$_2$ flakes in a more controlled way by using repetitive transfer processes. For this, we first prepare a suitable flake on top of a silicon substrate, and a second flake on top of PDMS. The second flake is then aligned and stamped onto the flake on the silicon substrate. This process allows us to create larger artificially stacked regions (limited by the size of the individual flakes), and typically gives optical access to the individual constituent layers in areas where the flakes do not overlap. However, in this process, the flake surfaces are exposed to ambient conditions for a substantial amount of time before the stacking operation is performed.
\subsection{Optical spectroscopy}
Raman spectroscopy measurements are performed at room temperature. For this, we use a microscope setup, in which a linearly polarized 532~nm cw laser is coupled into a 100x (1~$\mu$m laser spot size)  microscope objective, which also collects the scattered light in backscattering geometry. The scattered light is recorded using a  grating spectrometer equipped with a Peltier-cooled charge-coupled device (CCD) sensor. For most measurements of the higher-frequency Raman modes in MoS$_2$ (E$^1_{2g}$  and A$_{1g}$), a longpass filter suppresses the elastically scattered laser light. For measurements of the lower-frequency shear mode, a set of three reflective volume Bragg grating filters is used to deflect the elastically scattered laser light. Additionally, a polarizer is placed in front of the spectrometer and oriented so that only scattered light with linear polarization orthogonal to the laser is transmitted. The use of the Bragg grating filters gives access to, both, Stokes and Anti-Stokes Raman signals in the same spectrum.
The sample is mounted on a motorized XY table and scanned under the microscope. Room-temperature PL measurements are performed using the same setup. For low-temperature PL measurements, the samples are mounted in vacuum on the cold finger of  a small He-flow cryostat, which can also be scanned under the microscope. In PL and Raman scanning experiments, full spectra are collected for each laser spot position defined on a square lattice.

In order to extract information from these spectra, an automated fitting routine is employed, which yields the integrated intensity, spectral position and full width at half maximum (FWHM) of the characteristic PL and Raman spectral features. To study valley polarization effects, near-resonant excitation is employed in the PL setup. For this, we employ a 633~nm cw Helium-Neon laser. This laser is circularly polarized by a quarter-wave plate and coupled into the microscope system. A longpass filter is used to suppress scattered laser light, and the circular polarization of the PL is analyzed using a second quarter-wave plate and a linear polarizer placed in front of the spectrometer.

Second-harmonic generation (SHG) measurements are  performed at room temperature. For these measurements, a mode-locked Ti:sapphire laser generating 80~fs pulses is utilized in conjunction with a microscope setup similar to the one described above. Laser pulses tuned to a center wavelength of 830~nm are coupled into a 100x  microscope objective, which also collects the SHG in backscattering geometry. The backscattered light is spectrally filtered by a 600~nm shortpass and analyzed in a grating spectrometer equipped with a liquid-nitrogen-cooled CCD sensor.

In order to determine the crystallographic orientation of different MoS$_2$ flakes, linearly polarized excitation is utilized, and only the SHG with polarization parallel to the excitation is detected. The sample is rotated beneath the microscope setup. In SHG mapping experiments,  circularly polarized excitation is used, and there is no polarization analysis of the SHG intensity. The sample is scanned under the microscope, and the total SHG intensity is recorded for each sample position defined on a square array.
\section{Results and discussion}
\begin{figure}
   \includegraphics[width= \linewidth]{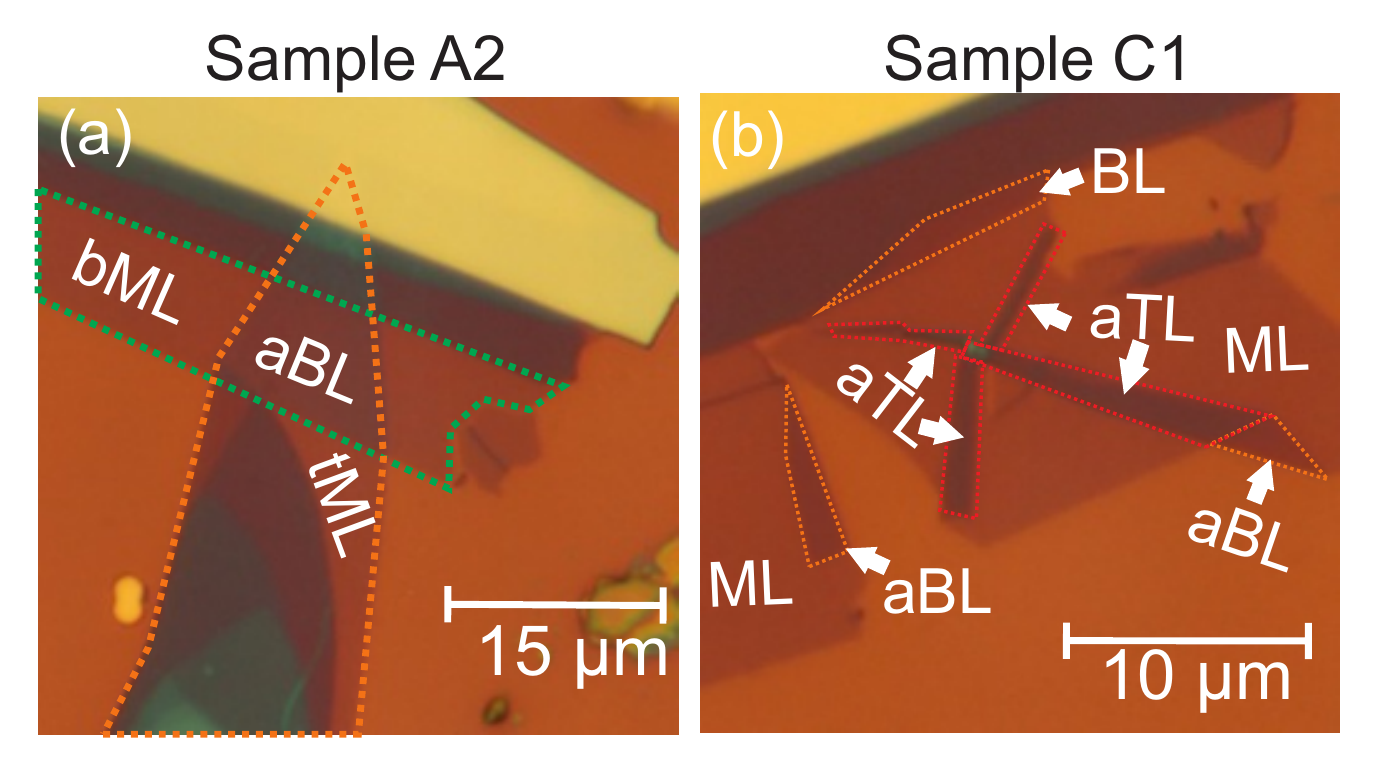}
   \caption{\textbf{Sample structures:} Optical microscopy images of samples A2(a) and C1(b). In (a), the outlines of the two monolayers (ML)  stacked on top of each other are marked by dashed lines (bottom and top ML are marked as bML and tML), and the artificial bilayer (aBL) region is indicated. In (b), the outlines of the 2H-stacked bilayer (BL), aBLs and artificial trilayers (aTL) created by twisting/folding are marked by dashed lines.}
   \label{Fig1}
\end{figure}

We will mainly discuss the results obtained on two different sample structures, which are depicted in figure~\ref{Fig1}. Sample A2 (Fig.~\ref{Fig1}(a)) consists of two MoS$_2$ flakes which were stacked on top of each other in subsequent transfer processes. Both flakes contain monolayers, and their overlapping region forms an artificial bilayer (aBL) whose stacking order differs from the naturally occurring 2H stacking. The top ML also covers few-layer areas of the bottom flake. Sample A3 (not shown) was prepared in a similar manner by stacking two flakes on top of each other. Sample C1 (Fig.~\ref{Fig1}(b)) consists of a single MoS$_2$ flake which was folded during the exfoliation process, so that in the center part of the large ML, 4 petal-like double folds are formed, which correspond to artificial trilayers (aTL). Additional folds are present at the edges of the ML, corresponding to aBLs.
\subsection{Raman spectroscopy}
\begin{figure}
   \includegraphics[width= \linewidth]{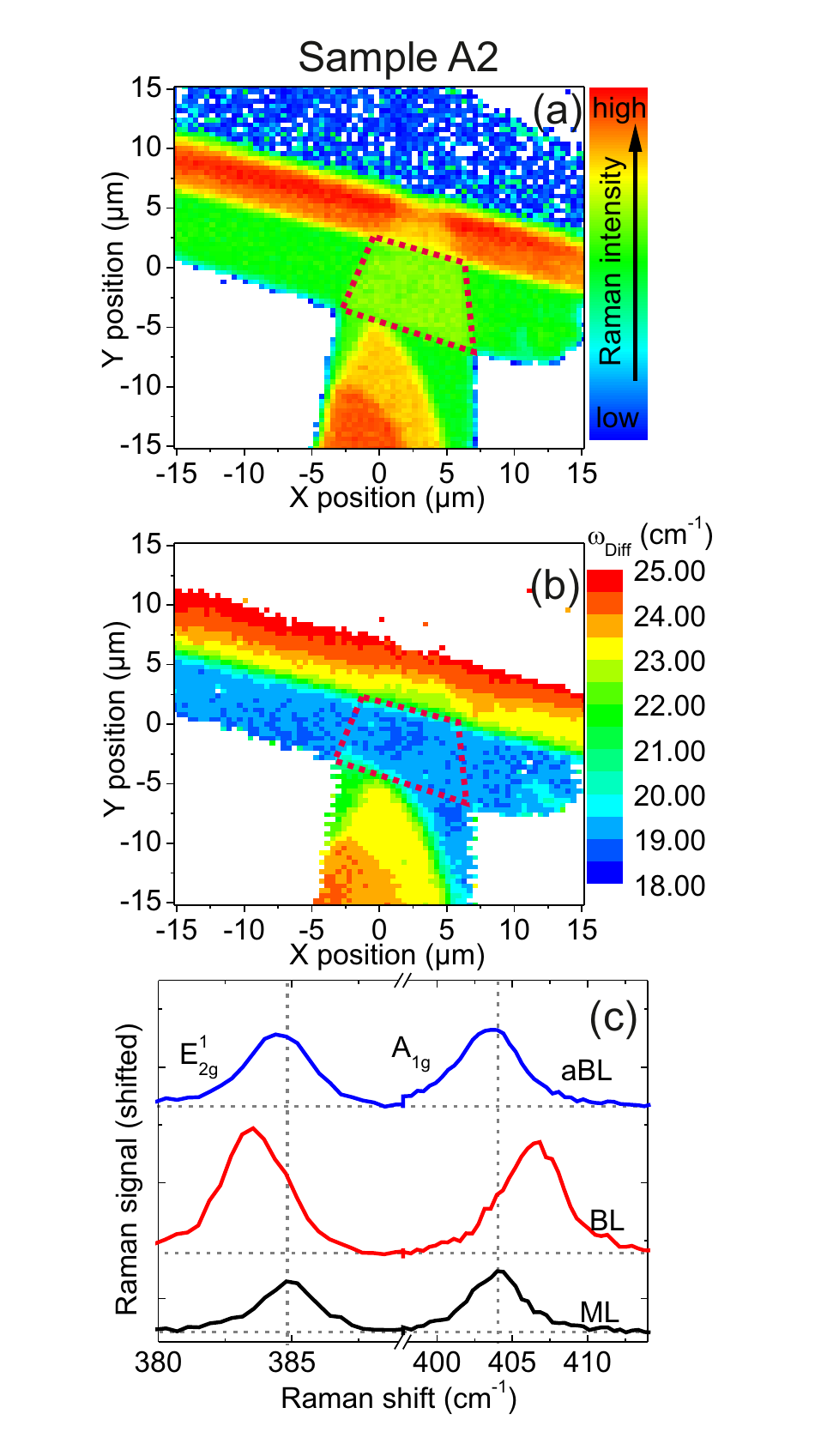}
   \caption{\textbf{Raman spectroscopy of sample A2:} (a)  False color plot of the $A_{1g}$ Raman mode intensity. (b) False color plot of the $A_{1g}-E^1_{2g}$ Raman mode frequency difference $\omega_{\textrm{Diff}}$.  The red dashed lines in (a) and (b) mark the outline of the aBL region. (c) Raman spectra of ML and aBL regions. A BL spectrum measured on sample C1 is included for comparison.}
   \label{Fig2_A2}
\end{figure}
\begin{figure}
   \includegraphics[width= \linewidth]{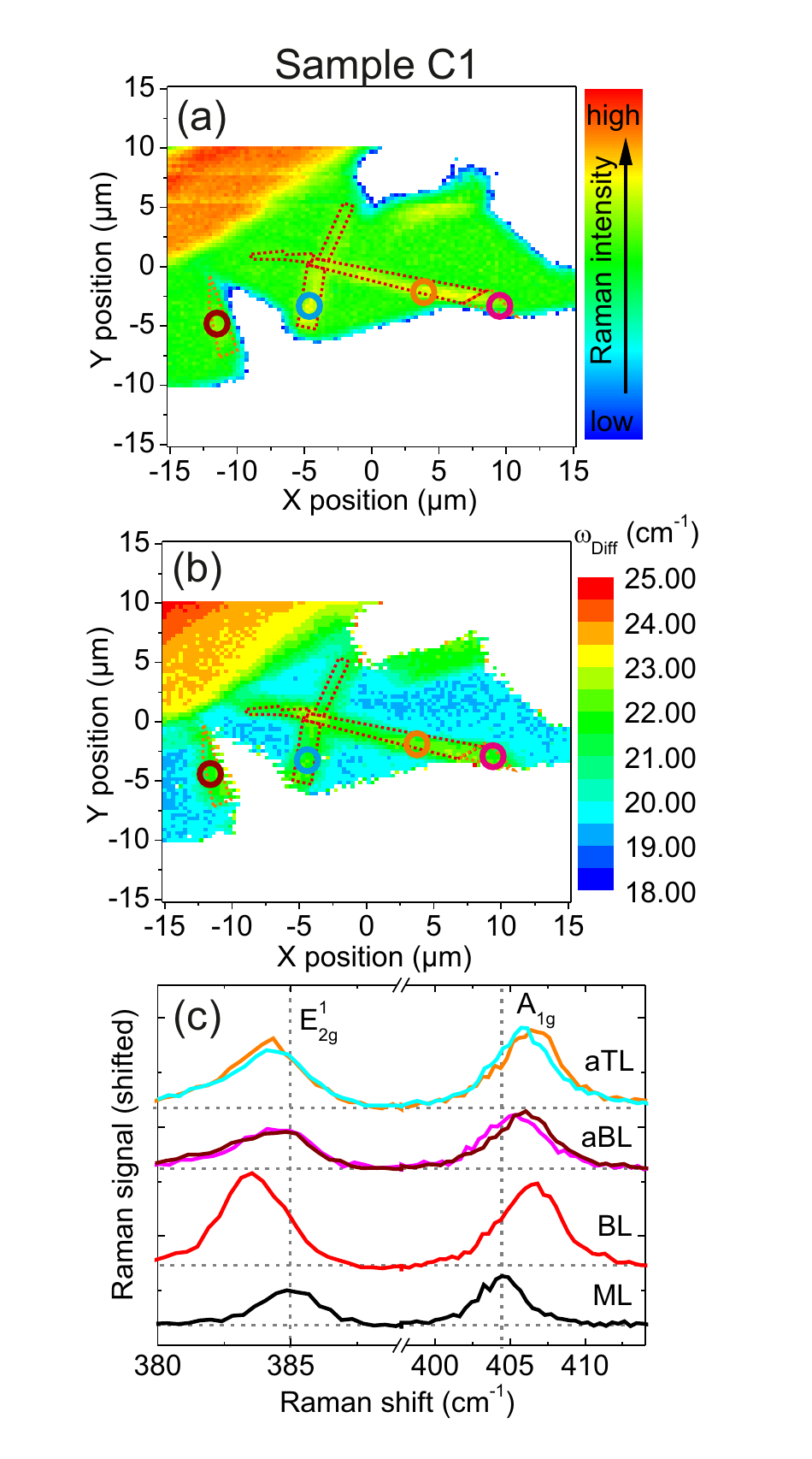}
   \caption{\textbf{Raman spectroscopy of sample C1:} (a)  False color plot of the $A_{1g}$ Raman mode intensity. (b) False color plot of the $A_{1g}-E^1_{2g}$ Raman mode frequency difference $\omega_{\textrm{Diff}}$. The red (aTL) and orange (aBL) dashed lines in (a) and (b) mark the outline of the artificially stacked regions. The colored circles in (a) and (b) mark the positions of the aBL and aTL Raman spectra shown in (c). (c) Raman spectra of ML, BL, aBL and aTL regions.}
   \label{Fig2_C1}
\end{figure}
First, we discuss  Raman spectroscopy of samples A2 and C1. Raman spectroscopy has been established as a fast, nondestructive means of characterization of MoS$_2$ flakes prepared from bulk, 2H-stacked mineral MoS$_2$. In these flakes, there is a pronounced dependence of the characteristic Raman mode frequencies on the number of layers. The A$_{1g}$ mode, which is an out-of-plane optical vibration of the S atoms, stiffens with additional layers due to increasing restoring forces acting on the S atoms. By contrast, the  E$^1_{2g}$ mode, which corresponds to an in-plane optical vibration of Mo and S atoms, anomalously softens with additional layers~\cite{Heinz_ACSNano10}. This softening is attributed to enhanced dielectric screening~\cite{Molina11} and next-nearest-neighbor interactions~\cite{Wirtz}. Therefore, the frequency difference $\omega_{\textrm{Diff}}= \omega_{A_{1g}}-\omega_{E^1_{2g}}$ between the two modes can be used as a fingerprint for the number of layers, at least for flakes of up to 4 layers, while for thicker layers the mode positions approach the bulk limit.
Additionally, the Raman intensity of the $A_{1g}$ and $E^1_{2g}$ modes may be used to quantify the number of layers in mechanically exfoliated flakes. The Raman intensity dependence on the number of layers is highly nonmonotonic due to substrate-induced and internal interference effects, and distinct intensity maxima are obtained for certain flake thickness values~\cite{Li_12}, with the first maximum at 4 layers.

We can compare these findings for 2H-stacked mineral MoS$_2$ to the results we obtain on artificially stacked  MoS$_2$. In Figures~\ref{Fig2_A2}(a) and \ref{Fig2_C1}(a), we plot false color maps of the Raman intensity of the $A_{1g}$ mode. The two false color plots clearly reproduce the topography of our samples, with the aBL and aTL layers showing higher Raman intensities than the ML parts. In sample A2, we also observe a decreased Raman intensity in the multilayer-region of the bottom flake that is covered by the top ML. In this multilayer region, the bottom flake consists of 4 and more layers, so that an additional layer decreases the Raman intensity due to interference effects. In sample C1, the aTL petals on the top and the left mostly have a width below the laser spot size of about 1~$\mu m$, so that they are less visible in the intensity map than the other two aTL petals due to spatial averaging.

We find remarkable differences between the samples in the Raman frequency difference maps shown in Figures~\ref{Fig2_A2}(b) and \ref{Fig2_C1}(b): in sample A2, the frequency difference between the $A_{1g}$ and $E^1_{2g}$ modes is identical for both ML flakes and the aBL region, so that the aBL region is not observable in the frequency difference map. Additionally, there is no effect of the top ML on the frequency difference in the multi-layer region of the bottom flake. By contrast, in sample C1, all aBL and aTL regions show a larger frequency difference than the ML, so that the frequency difference map clearly reproduces the topography of the flake. We  directly compare the Raman spectra for different regions in samples A2 and C1 in Figures~\ref{Fig2_A2}(c) and \ref{Fig2_C1}(c). For sample A2, we see that both Raman modes for the aBL are slightly redshifted (by less than 0.5~cm$^{-1}$) as compared to the ML, in striking contrast to the modes of a 2H-stacked BL (in Figure~\ref{Fig2_A2}(c), a spectrum of a 2H-stacked BL measured on sample C1 is included for direct comparison), where a blueshift of the $A_{1g}$ by about 2.5~cm$^{-1}$ and a redshift of the $E^1_{2g}$ by about 1~cm$^{-1}$ occur. This is a clear indication that interlayer coupling in the aBL of sample A2 is significantly reduced as compared to a 2H-stacked BL. The small redshift we observe for both modes in the aBL most likely stems from dielectric screening, which is less sensitive to the relative crystallographic orientation of adjacent layers than, e.g., van der Waals force constants between atoms.
In sample C1, we find that the aBL and aTL regions show Raman mode shifts that are qualitatively similar to 2H-stacked MoS$_2$ - however, direct comparison to a 2H-stacked BL shows that, both, the blueshift of the $A_{1g}$ mode and the redshift of the $E^1_{2g}$ mode are less pronounced, even for the aTL regions, indicating that interlayer coupling in this sample is still reduced compared to regular 2H crystals.

Remarkably, we can also observe clear differences in the sensitivity of the Raman modes on the interlayer coupling: while the $E^1_{2g}$ mode positions for two different aBL and aTL regions (the measurement positions are indicated by the colored circles in the false color plots in Fig.~\ref{Fig2_C1}(a) and (b)) match, the $A_{1g}$ mode positions for these regions, which have different relative crystallographic orientations, differ by about 1~cm$^{-1}$. This finding supports our interpretation that the redshift of the $E^1_{2g}$ mode in the artificially stacked layers is due to dielectric screening, which is less sensitive to the stacking angle, while the out-of-plane $A_{1g}$ mode frequency is sensitive to nearest-neighbor-interactions between adjacent layers. The frequency difference of about 22~cm$^{-1}$ for the aBL regions in sample C1, and the larger dependency of the $A_{1g}$ mode on stacking angle, are in qualitative agreement with Raman spectroscopy characterization of aBLs reported by other groups. However, we should note the different source materials: while our aBLs are prepared from exfoliated MoS$_2$ flakes, the other groups~\cite{Hone_Twist_NanoLett14,Wang_Twist_NatComm14,Dresselhaus_twisted_NL} utilized CVD-grown MoS$_2$ triangular islands. Direct comparison of the Raman modes in exfoliated ML flakes and CVD-grown, triangular MLs show an increased frequency difference for the CVD-grown material~\cite{Plechinger_SST14}, most likely due to an increased interaction with the SiO$_2$ substrate.
\begin{figure}
   \includegraphics[width= \linewidth]{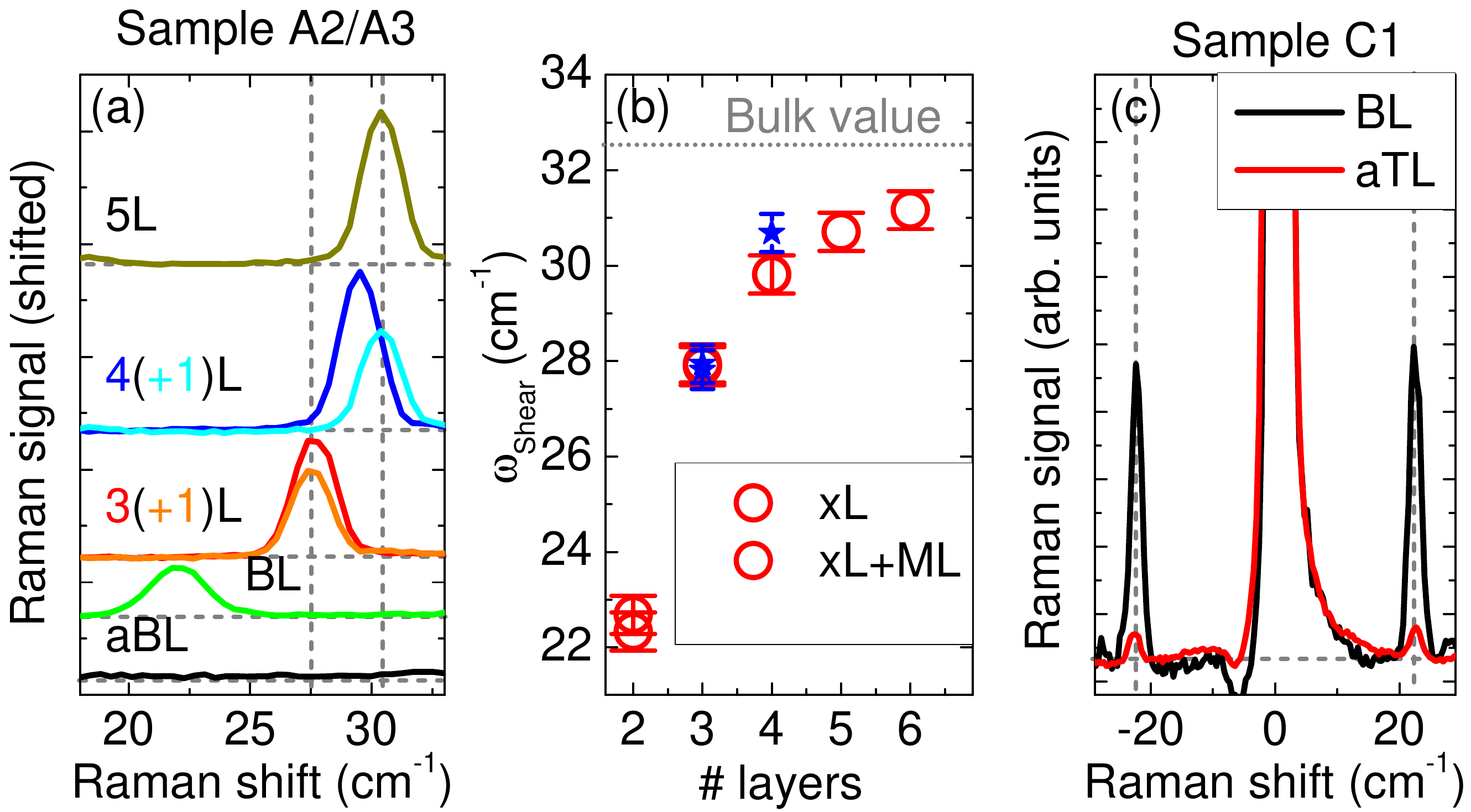}
   \caption{\textbf{Low-frequency Raman spectroscopy:}(a)  Low-frequency Raman spectra measured at different positions on samples A2 and A3. The orange (3L+ML) and light blue traces (4L+ML) indicate measurements on layers of known thickness covered with an additional monolayer. (b) Shear mode frequency as a function of the number of layers. Circles indicate measurements on naturally stacked layers, stars measurements on regions covered with an additional monolayer. (c) Raman spectra of naturally stacked bilayer and aTL region in sample C1.}
   \label{Fig3}
\end{figure}

We now to turn to low-frequency Raman spectroscopy measurements, which give access to the rigid-layer vibrational modes that are generic to layered materials. These vibrational modes have low frequencies due to the weak van der Waals forces between individual layers. Naturally, these modes are absent in MLs, and they also have a characteristic dependency on the number of layers~\cite{Ferrari_NatMat12,plechinger:101906,Ferrari_ShearMos,Dresselhaus_Shear}. The interlayer Raman modes that yield the largest observable signal in MoS$_2$ are the highest-frequency interlayer shear mode and  the lowest-frequency interlayer breathing (or compression) mode that are allowed for a certain number of layers. While the shear mode shifts to higher frequencies with the number of layers, the breathing mode shifts to lower frequencies.

 In our low-frequency Raman measurements, we detect  signals with polarization perpendicular to that of the laser, so that due to polarization selection rules, we are sensitive to the interlayer shear mode, only. Figure~\ref{Fig3} summarizes the results of our measurements: in samples A2 and A3, MoS$_2$ MLs are stacked on top of MoS$_2$ flakes containing ML and multilayer regions, so that, both, aBLs and few-layer MoS$_2$ with an additional ML on top are available for study. While 2-H-stacked bilayers show a clear shear mode signal, we do not observe the shear mode in the aBL regions of either sample in the accessible frequency range above 10~cm$^{-1}$, indicating a weak interlayer coupling with reduced van der Waals interaction between adjacent layers. In multilayer regions covered with an additional ML, we mostly see that the additional ML only leads to a slight reduction of the Raman signal amplitude of the shear mode, while its position typically corresponds to that of a multilayer without the additional ML, as seen in the red and orange spectra in Figure~\ref{Fig3}(a). A deviation from this behavior is only observed at one position in sample A2 where a four-layer region is covered with an additional ML. Here, a shear mode signal corresponding to the  frequency for 5 layers is observed (light-blue spectrum in  Figure~\ref{Fig3}(a)), albeit with a reduced intensity compared to an adjacent 2H-stacked 5-layer region (dark-yellow spectrum). All results of measurements on samples A2 and A3 are summarized in ~\ref{Fig3}(b), multiple datapoints in this graph indicate measurements on different regions with the same number of layers. With the exception discussed above, the shear mode frequencies observed are in good agreement with previous reports for 2H-stacked MoS$_2$~\cite{plechinger:101906,Ferrari_ShearMos,Dresselhaus_Shear}.

In sample C1, in which the previous Raman measurements indicated a larger interlayer coupling than in sample A2, almost all of the aBL and aTL regions are devoid of shear mode signals. However, in a part of the aTL region of the left petal, we find a weak signal corresponding to the mode frequency of a 2-H-stacked bilayer (red spectrum in  Figure~\ref{Fig3}(c)), indicating a local change of the interlayer coupling to larger values. While the aTL region in which we observe this shear mode signal is significantly larger than the laser spot size, the weak signal indicates that the shear mode only develops in small patches of the aTL. In the photoluminescence measurements discussed below, we do not observe effects related to locally enhanced interlayer coupling in this region.
The low-frequency Raman spectroscopy results indicate that the shear mode is highly sensitive to the interlayer coupling, but unsuitable to determine a gradual decoupling of adjacent layers, as it is not observable under these conditions.
\subsection{Second harmonic generation microscopy}
\begin{figure}
   \includegraphics[width=1.05 \linewidth]{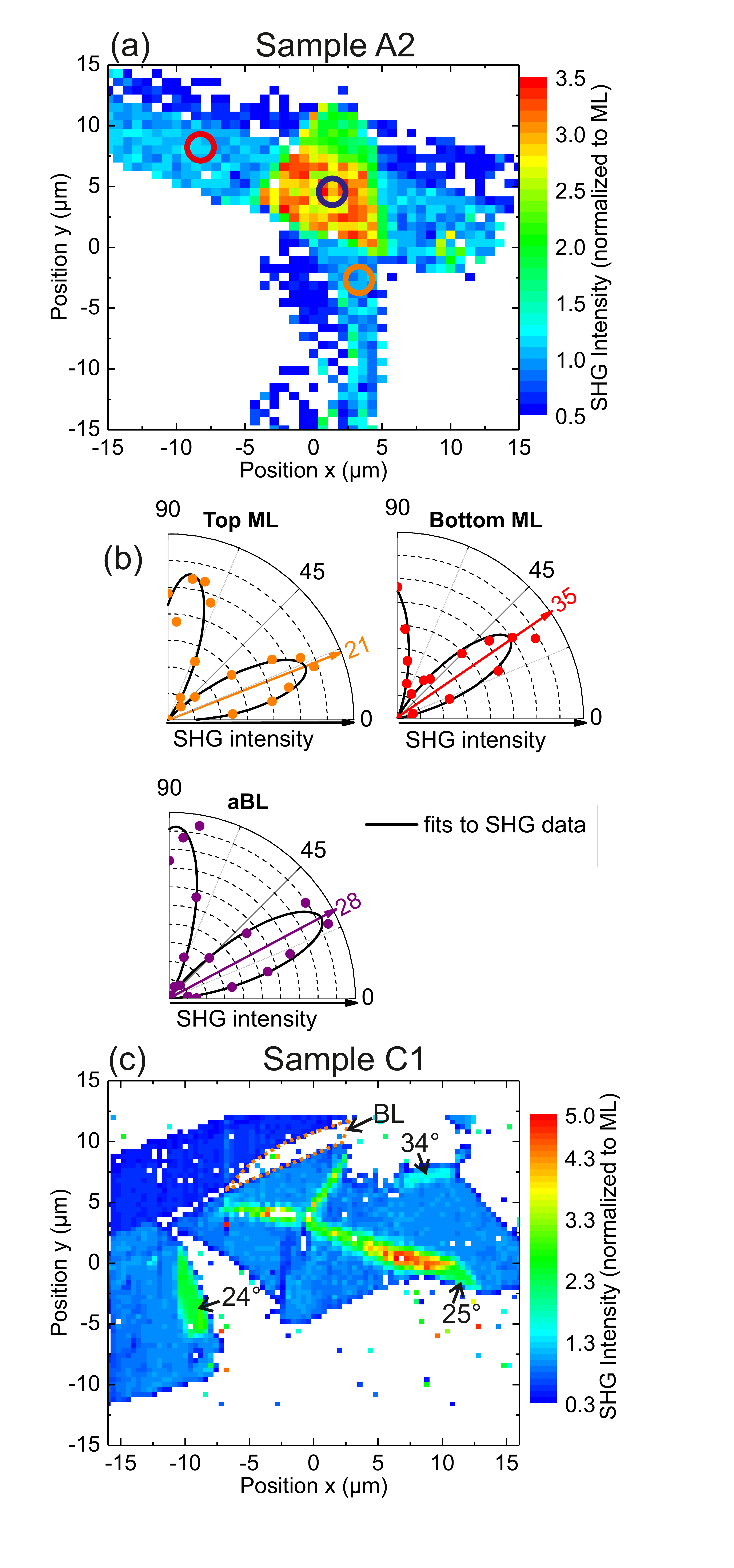}
   \caption{\textbf{Second-harmonic-generation microscopy:} (a)  False color plot of the SHG intensity under circularly polarized excitation for sample A2. The colored circles indicate the regions where the polarization-dependent measurements shown in (b) were performed. (b) Polar plots of the SHG intensity dependence on polarization under linearly polarized excitation for the two ML and the aBL regions on sample A2. Solid lines indicate fits to the data. (c) False color plot of the SHG intensity under circularly polarized excitation for sample C1. The stacking angles determined from the SHG intensity for three different aBL regions are indicated by the black arrows. The dashed orange outline marks the 2H-stacked BL region of sample C1 which does not show any SHG intensity.}
   \label{Fig4}
\end{figure}
We now turn to the discussion of SHG microscopy measurements. This technique~\cite{Heinz_SHG13,Malard_SHG13} has emerged as a highly useful tool to determine the crystallographic orientation of dichalcogenide monolayers, to clearly separate odd- and even-numbered layer regions in 2H-stacked dichalcogenides, and to study the stacking angle in twisted bilayers~\cite{Chang_SHG14}. SHG is only allowed in crystal structures without an inversion center. In few-layer 2H-stacked flakes of MoS$_2$ and related dichalcogenides like WS$_2$, WSe$_2$ and MoSe$_2$, the crystallographic point group, and the corresponding presence of an inversion center, depends on the number of layers: structures with an odd number of layers belong to the D$_{3h}$ point group, which lacks an inversion center, while structures with an even number of layers belong to the D$_{3d}$ point group, which contains an inversion center. Therefore, SHG is suppressed in even-numbered 2H-stacked flakes. In odd-numbered layers, polarization-dependent SHG measurements allow to determine the crystallographic orientation: the electric field amplitude $E_{2\omega \|}$ of the second harmonic that is polarized parallel to the fundamental laser field amplitude $E_{\omega}$  depends on the angle $\phi$ between the laser field polarization and the armchair direction of the crystal lattice: $E_{2\omega \|}\sim \cos(3\phi)$. The resulting SHG intensity $I_{2\omega \|}= E^2_{2\omega \|}  \sim \cos^2(3\phi)$ has six-fold symmetry~\cite{Jiang_stacked_Valley_NatNano14}. The SHG component $I_{2\omega \perp}$ polarized perpendicular to the fundamental laser field is given by $I_{2\omega \perp} \sim \sin^2(3\phi)$, so that the total SHG intensity is independent of $\phi$. In odd-numbered 2H-stacked flakes, the total SHG intensity drops with increasing number of layers due to interlayer coupling and resulting changes in the band structure~\cite{Heinz_SHG13}.

By contrast, in artificial bilayer stacks, in which the individual layers are decoupled, the SHG intensity varies with the relative stacking angle due to interference between the SHG fields of the monolayers~\cite{Chang_SHG14}. The total SHG intensity for an artificial bilayer is given by
 \begin{equation}
 I_{2\omega}(\varphi)= I_1+I_2+2\sqrt{I_1 I_2}\cos(3\varphi),
    \label{Stacking}
 \end{equation}
where $I_1$ and $I_2$ are the total SHG intensities for the individual MLs and $\varphi$ is the stacking angle, e.g., the angle between two armchair directions in the adjacent layers. Thus, for equal  $I_1$ and $I_2$, $I_{2\omega}$ may vary between zero and $4 I_1$. It should be noted that even in aBLs which show zero SHG intensity, the alignment of the layers may differ from the 2H stacking, e.g., due to an in-plane translation of the layers relative to each other, or an increased interlayer spacing.
Figure~\ref{Fig4}(a) shows a false color plot of the total SHG intensity of sample A2 under circularly polarized excitation. The color scale has been normalized to the SHG intensity of the ML regions. Averaging the SHG intensity in the aBL region we find that it is about 3 times larger than in the ML, indicating partially constructive interference of the individual ML SHG fields. Using equation~\ref{Stacking} and utilizing that $I_1=I_2$, as seen in Figure~\ref{Fig4}(a), we can estimate the stacking angle to be $\varphi \approx 20$~degrees.

A more precise measurement of the stacking angle is possible by performing angle-dependent SHG intensity measurements on the individual MLs and the aBL, using linearly polarized excitation and detection of $I_{2\omega \|}$ with an analyzer in the beam path. The results are depicted in the polar plots in Figure~\ref{Fig4}(b). For each  ML and the aBL, $I_{2\omega \|}$ has a six-fold symmetry, so that angle-dependent measurements spanning a range of 90~degrees are sufficient to determine the orientation of an armchair direction in the MLs with respect to  horizontal polarization of the incident electric field, corresponding to $\phi=0$. By fitting the measured SHG intensities using $I_{2\omega \|}= C \cdot \cos^2(3\phi+\phi_0)$, we find  armchair orientations of 35~degrees (bottom ML) and 21~degrees (top ML), yielding a relative stacking angle of $\varphi = 14$~degrees. As expected, the maximum SHG intensity for the aBL is observed for an angle of 28~degrees, which is the average of the angles observed for the two MLs.

Figure~\ref{Fig4}(c) shows a false color plot of the total SHG intensity of sample C1 under circularly polarized excitation.  The color scale has been normalized to the SHG intensity of the ML region. We clearly see that the total SHG intensity for the different aBL regions is different, but always significantly larger than that of the ML. By contrast, the 2H-stacked BL region of the sample (marked by the orange outline) does not show any SHG intensity due to the inversion symmetry of the naturally stacked bilayer. We can again estimate the relative stacking angles for the aBL regions using equation~\ref{Stacking}. The stacking angles are given in Figure~\ref{Fig4}(c). For three of the four petals formed by the aTLs, we also observe increased SHG intensity, while the lower left petal aTL shows a reduced intensity. Given that these regions consist of three layers with two independent stacking angles between adjacent MLs, we cannot determine the stacking angles from the total SHG intensity, only. In contrast to sample A2, the crystallographic orientation of the constituent MLs of the artificially stacked regions in sample C1 cannot be independently established by polarization-resolved SHG measurements due to the folding.
\subsection{Photoluminescence spectroscopy}
\begin{figure}
   \includegraphics[width= 1.05 \linewidth]{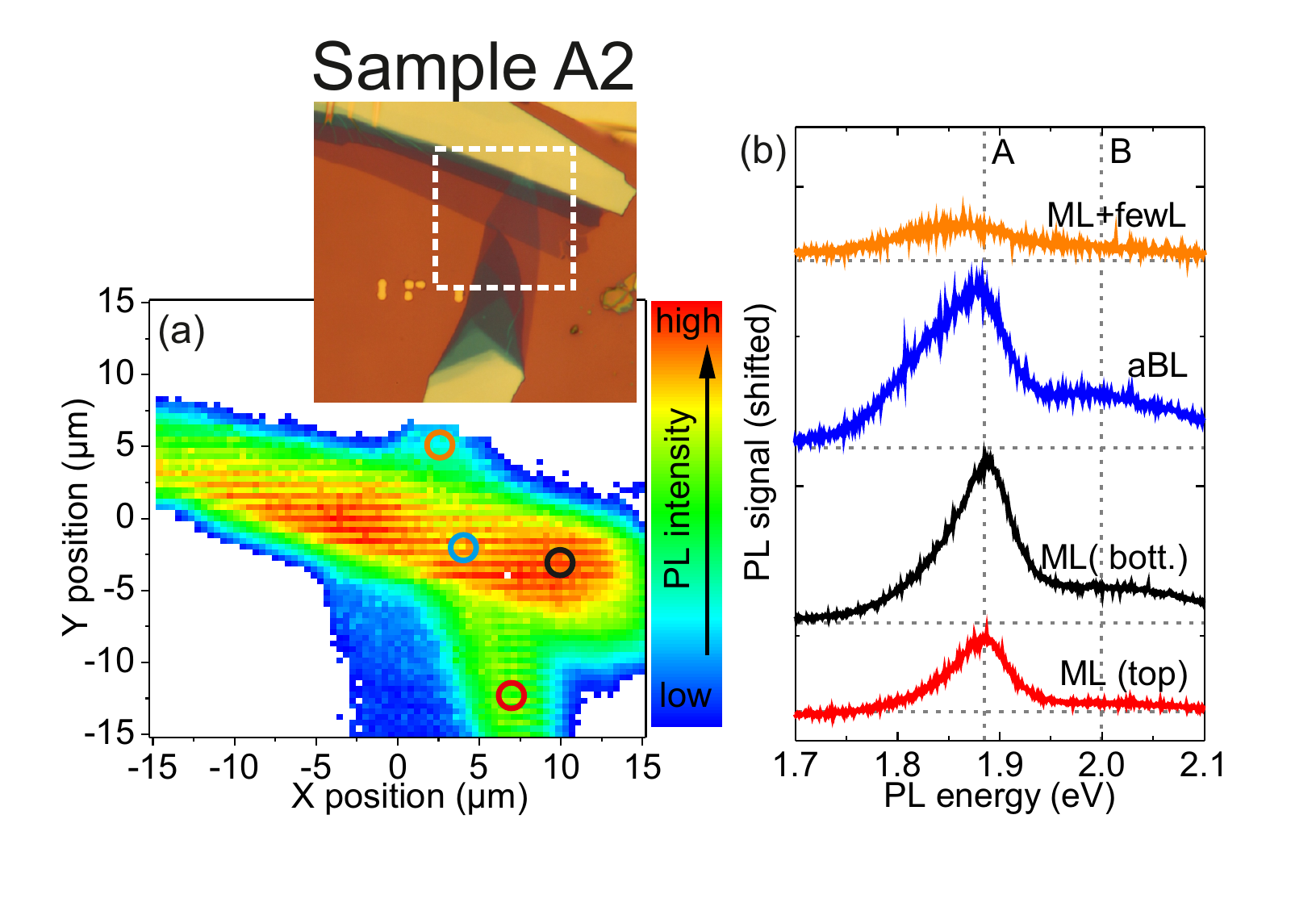}
   \caption{\textbf{Room-temperature PL measurements on sample A2:} (a) False color plot of the A exciton PL intensity. The colored circles mark the positions of the PL spectra shown in (b). The microscope image in the inset shows the PL scan region marked by the dashed white outline. (b) PL spectra of top and bottom ML,  aBL, and ML on few-layer MoS$_2$ regions.}
   \label{Fig5_A2}
\end{figure}
\begin{figure}
   \includegraphics[width=  1.05  \linewidth]{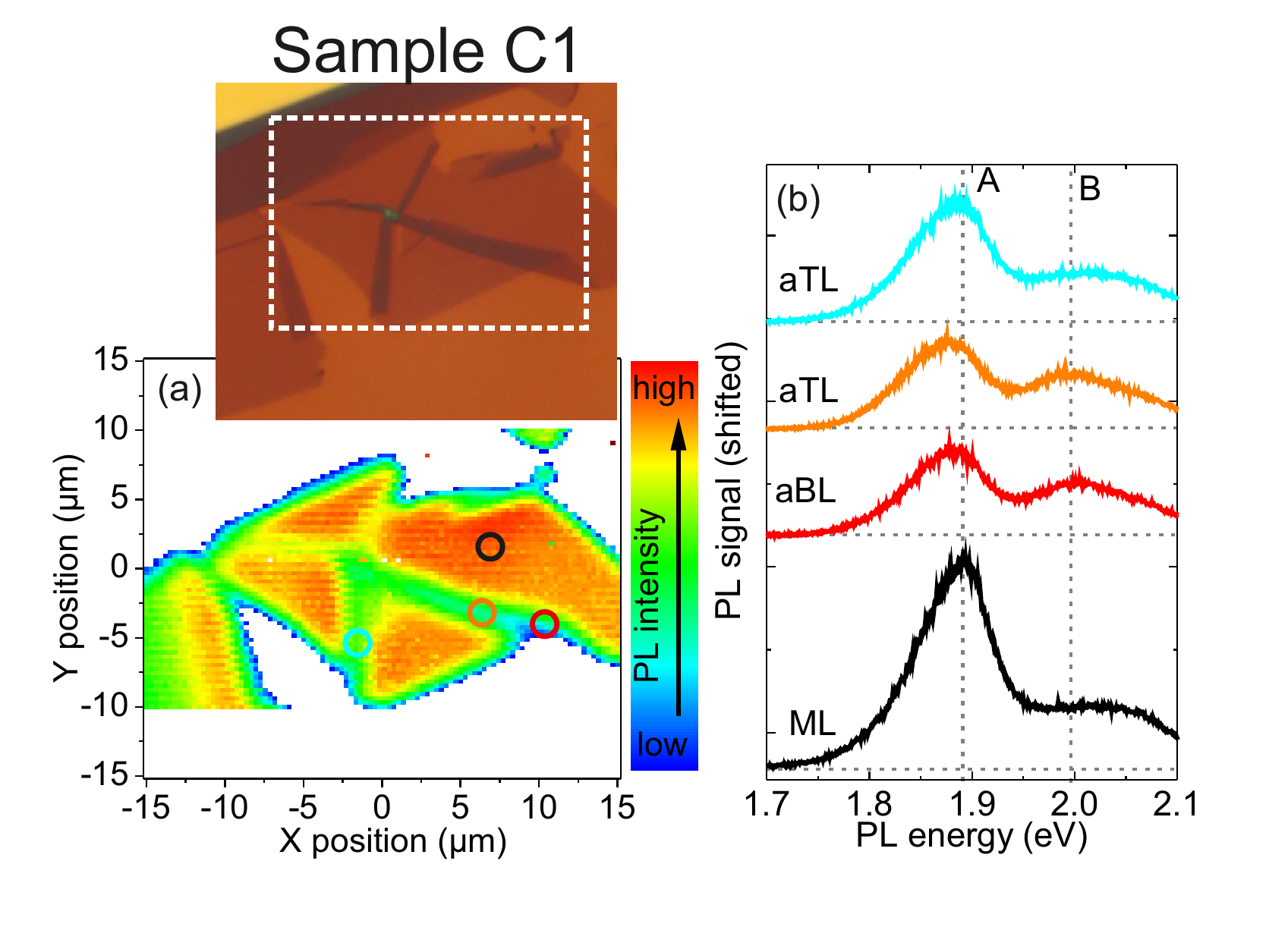}
   \caption{\textbf{Room-temperature PL measurements on sample C1:} (a) False color plot of the A exciton PL intensity. The colored circles mark the positions of the PL spectra shown in (b). The microscope image in the inset shows the PL scan region marked by the dashed white outline. (b) PL spectra of ML,  aBL, and aTL regions.}
   \label{Fig5_C1}
\end{figure}
Next, we discuss photoluminescence measurements on our samples. In recent years, PL has been used to study the transition from an indirect to a direct band gap in MoS$_2$ monolayers~\cite{Heinz_PRL10, Splen_Nano10}, and to investigate valley~\cite{Yao12,Heinz12,kioseoglou:221907} and (charged) exciton~\cite{Heinz_Trions, Ross_13} physics.  We first present room-temperature PL results. Figures~\ref{Fig5_A2}(a) and \ref{Fig5_C1}(a) show false color plots of the A exciton PL intensity in samples A2 and C1. In sample A2, we find that the two ML regions have different PL intensities, with the bottom ML PL being about twice as intense (black and red spectra in Figure~\ref{Fig5_A2}(b)). Such variations in PL intensities between individual MoS$_2$ flakes are quite common due to the mineral bulk source material. Remarkably, the aBL region (blue spectrum in Figure~\ref{Fig5_A2}(b)) shows an A exciton PL intensity that is similar to the bottom ML, with a slight broadening and redshift of the emission by about 10~meV. This observation is in striking contrast to previous reports on aBLs~\cite{Hone_Twist_NanoLett14,Dresselhaus_twisted_NL,Jiang_stacked_Valley_NatNano14}, where a pronounced suppression of the A exciton emission was seen. In two of these studies~\cite{Hone_Twist_NanoLett14,Jiang_stacked_Valley_NatNano14}, this reduction was combined with the appearance of a lower-energy PL peak, indicating a changeover in the aBLs to an indirect band gap. Such a low-energy peak is absent in our samples. Therefore, we can infer that the MLs forming our aBL structure are electronically decoupled and retain the direct-gap band structure of an isolated ML. The most likely reason for this decoupling is the presence of adsorbates in between the individual layers, which increase the interlayer distance in the artificial structure. We find, however, that the aBL PL emission in our sample cannot simply be described by a sum of the two ML PL spectra, as we would expect for decoupled layers.
In the few-layer (fewL) region covered by the top ML, a PL signal which stems from the ML is still observable (orange spectrum in Figure~\ref{Fig5_A2}(b)), but  also redshifted and significantly weaker than that of the top ML on the bare SiO$_2$ substrate.

A  similar behavior is observed in sample C1, see Figure~\ref{Fig5_C1}: here, the PL intensity emitted from aBL (red spectrum in Figure~\ref{Fig5_C1}(b)) and aTL (blue and orange spectra) regions is also redshifted by about 10~meV, but even weaker than that of the ML (black spectrum).
To explain the changes of the PL emission in our artificially stacked regions, we need to consider resonant  energy transfer (RET) processes between the individual layers. Such a transfer process presents an additional decay channel for excitons, which may either recombine radiatively, emitting PL, decay nonradiatively within the ML, or transfer their energy to the adjacent layer via RET. Given that the quantum yield for ML MoS$_2$ is well below 1~percent~\cite{Heinz_PRL10}, almost all of the energy transferred between layers via RET will be lost due to nonradiative processes. Thus, RET acts as an additional luminescence-quenching mechanism in the artificial structures, which can be  described by an additional nonradiative decay rate, which needs to be added to the nonradiative decay rate within an individual layer. Naturally, this mechanism does not influence the instantaneous scattering processes studied by Raman and SHG microscopy. Nonradiative energy transfer into two-dimensional crystal structures, especially graphene, has been studied intensely in recent years~\cite{Stauber11, Koppens13, Ajayi14}. Due to the large, in-plane dipole moment~\cite{Schuller13} of excitons in ML MoS$_2$, nonradiative energy transfer between MLs can be very efficient, but being mediated by dipole-dipole interaction, it has a very strong dependency on the distance $d$ between energy donor and accepting layer, which was found to scale like $d^{-2.5}$ for ML MoS$_2$~\cite{Prins14}.

This strong distance dependency may explain the qualitative differences we observe between samples A2 and C1: in sample C1, we see  more pronounced changes to the Raman spectra in the aBL and aTL regions than in sample A2, which indicates larger interlayer coupling due to a smaller interlayer distance. Hence, we may expect RET between individual MLs to be more efficient in sample C1, yielding a larger nonradiative recombination rate. This leads to an enhanced suppression of the PL emission, as observed in our measurements. The redshift of the PL emission that we observe may also be attributed to the RET mechanism. RET will preferentially occur from states that spectrally match the absorption maximum of the adjacent layer. The PL emission of ML MoS$_2$ flakes is inhomogeneously broadened, in part due to surface adsorbates~\cite{Plechinger12}, and Stokes-shifted to lower energy with respect to the absorption maximum. Thus, RET will predominantly deplete states corresponding to the higher-energy part of the PL emission, so that an effective redshift occurs.
\begin{figure}
   \includegraphics[width= 1.05 \linewidth]{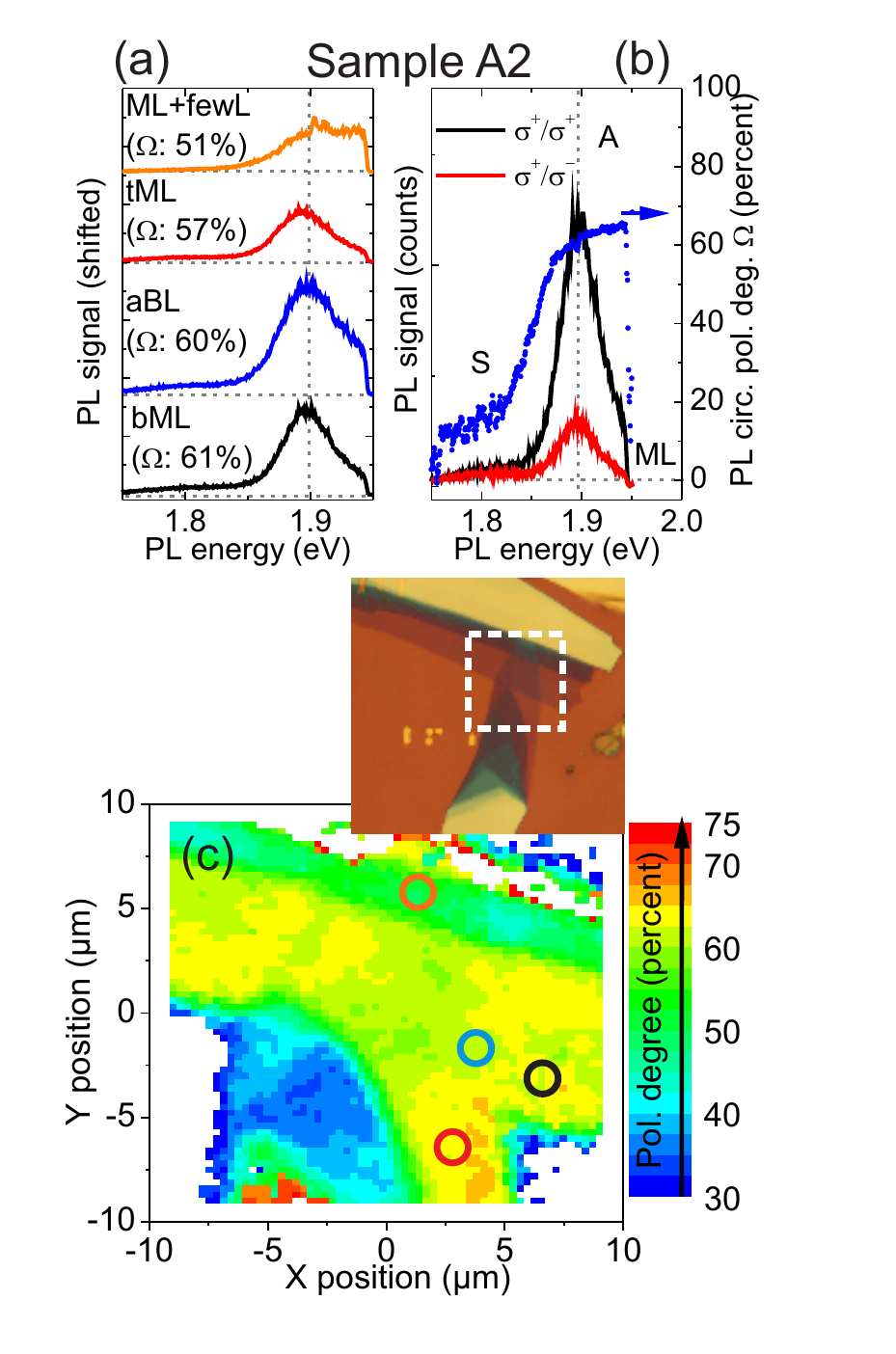}
   \caption{\textbf{Low-temperature PL measurements using near-resonant excitation on sample A2:} (a) PL spectra of top and bottom ML, aBL and ML   on few-layer MoS$_2$ regions. Each spectrum is the sum of the spectra collected under co- and contra-circular excitation/detection.  The spectrally averaged PL polarization degree is given next to the spectra. (b) helicity-resolved PL spectra of bottom ML region and resulting PL polarization degree. (c)  False color plot of the PL polarization degree. The colored circles mark the positions of the PL spectra shown in (a). The microscope image in the inset shows the PL scan region marked by the dashed white outline.}
   \label{Fig6_A2}
\end{figure}
\begin{figure}
   \includegraphics[width= \linewidth]{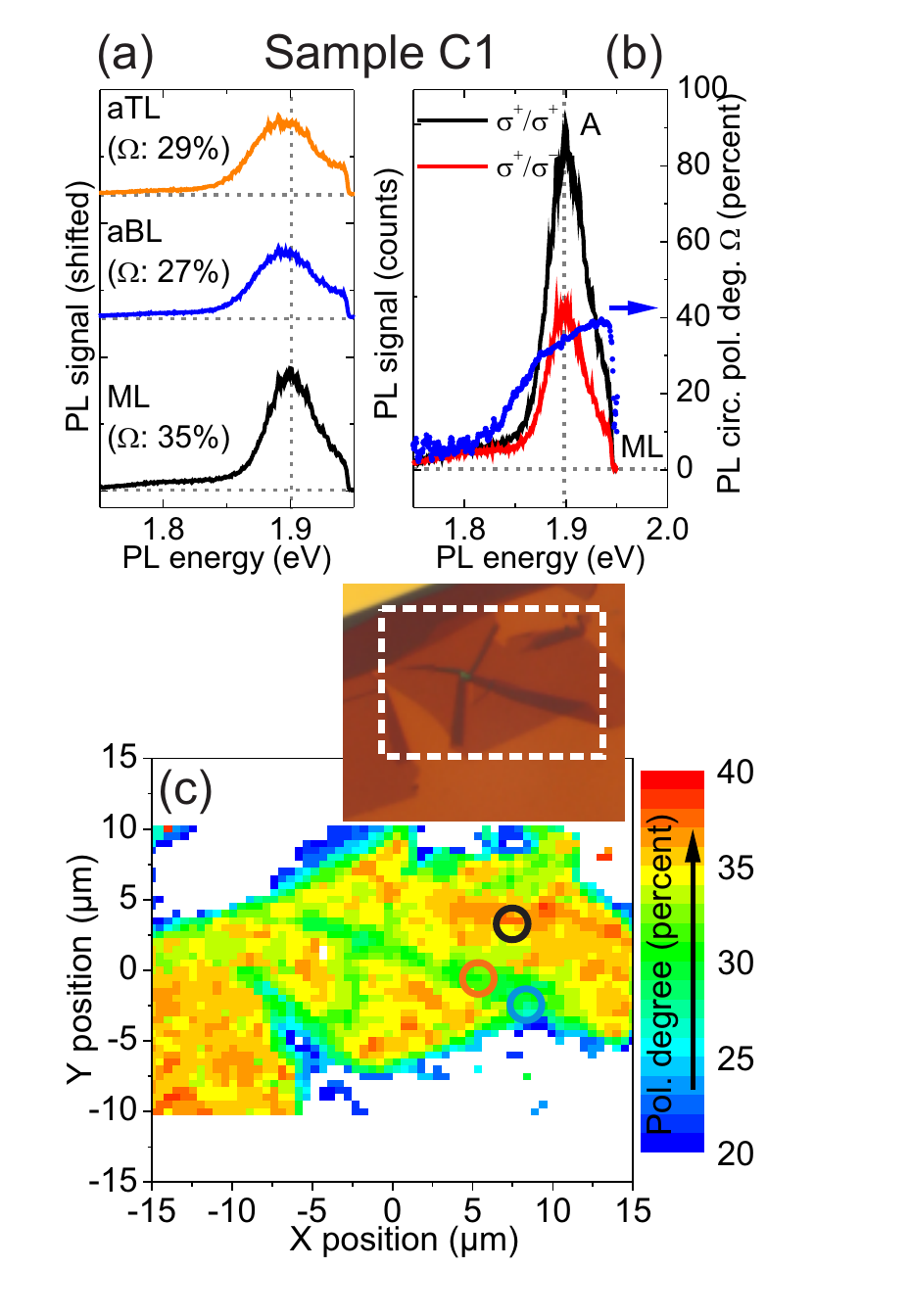}
   \caption{\textbf{Low-temperature PL measurements using near-resonant excitation on sample C1:} (a) PL spectra of ML, aBL and aTL regions. Each spectrum is the sum of the spectra collected under co- and contra-circular excitation/detection.  The spectrally averaged PL polarization degree is given next to the spectra. (b) helicity-resolved PL spectra of  ML region and resulting PL polarization degree. (c)  False color plot of the PL polarization degree.  The colored circles mark the positions of the PL spectra shown in (a). The microscope image in the inset shows the PL scan region marked by the dashed white outline.}
      \label{Fig6_C1}
\end{figure}

Low-temperature PL measurements confirm the presence of the RET mechanism. As Figure~\ref{Fig6_A2}(a) shows, there are qualitative changes in the PL emission of the different regions in sample A2: at liquid-helium temperature, the PL intensity of the aBL region (blue spectrum in Figure~\ref{Fig6_A2}(a)) is higher than that of the bottom ML (black spectrum), and there is no discernible redshift. For the ML region on top of the few-layer MoS$_2$ (orange spectrum), the PL intensity is not reduced compared to that of the top ML on the bare SiO$_2$ substrate (red spectrum), in stark contrast to the PL results obtained at room temperature in this region.

In sample C1 (see Fig.~\ref{Fig6_C1}(a)), a similar, but less pronounced trend is observable, with reduced redshifts and increased PL intensity in the aBL (blue spectrum in Figure~\ref{Fig6_C1}(a) and aTL(orange spectrum) regions compared to the ML region (black spectrum). The changes observed in the PL emission of the artificially stacked regions in our samples indicate that the RET mechanism is effectively suppressed at low temperatures. This suppression is due to the changing photocarrier dynamics: time-resolved PL measurements show that at low temperatures, the PL decays on the few-ps timescale, while  a longer-lived component develops with increasing temperature~\cite{Korn_APL11,Lagarde14}, most likely due to exciton-phonon scattering processes. Thus, reducing the temperature will increase the radiative recombination rate, while the competing RET rate, which is determined by the interlayer coupling, remains constant, so that the RET process is effectively suppressed, leading to the observed changes in PL intensity and position. In sample C1, which shows a stronger interlayer coupling, the RET rate is larger, so that temperature-induced changes of the photocarrier recombination dynamics have a smaller impact.

Finally, we discuss the valley polarization in our samples.
In our samples A2 and C1, we find that already for the ML regions, the observed circular polarization degree varies substantially, as seen in Figures~\ref{Fig6_A2}(a),(b) and \ref{Fig6_C1}(a),(b). For sample A2, we find  average values of about 61~percent and 57~percent for the bottom and top MLs, respectively, while for sample C1, the value is substantially lower, only 35~percent. In both samples, the A exciton transition energy is similar, so that we can exclude depolarization effects due to different excess energies provided by the fixed laser excitation~\cite{kioseoglou:221907}. Therefore, we have to attribute these  variations from flake to flake to different concentrations of crystal defects and ionized impurities, which influence momentum scattering rates. While the mechanisms for valley depolarization are currently under intense discussion~\cite{Dery_Relax,MWWu14,Glazov_14,Wu_EY_DP_MOS}, momentum scattering at defects is a possible valley dephasing channel. In the aBL region of sample A2, we do not observe any apparent reduction of the valley polarization compared to the MLs, as indicated in the false color map in Figure~\ref{Fig6_A2}(c).

By contrast, in sample C1, all  aBL and aTL regions show a clear reduction of the valley polarization by about 7~percent, which is near-homogenous for all artificial regions, see Figure~\ref{Fig6_C1}(a) and (c). Given that the SHG mapping for sample C1 clearly demonstrates different stacking angles for the different aBL and aTL regions, and that Raman spectroscopy indicates different interlayer coupling, we can exclude a systematic influence of these parameters on the valley polarization in our sample. One possible explanation for the reduced valley polarization in the aBL and aTL regions is an increased effective density of scattering centers in these folded regions: an ionized impurity present in the flake may also provide a scattering center in the adjacent layer within the folded region, as its stray field modifies the local potential. Therefore,  the density of scattering centers due to ionized impurities effectively doubles in the folded region compared to the isolated monolayer. In sample A2, this effect is not observable for two reasons: based on the large circular polarization degree observed in the individual monolayer regions, we may assume that the defect concentration is lower in this sample. Additionally, we can infer from our Raman spectroscopy measurements that the interlayer distance is larger in the aBL region of sample A2 than in the aBL and aTL regions of sample C1. Hence, the effect of an ionized impurity in one layer on the local potential in an adjacent layer will be reduced.
\section{Conclusion}
In summary, we have used optical spectroscopy to study the electronic and vibrational coupling in artificial MoS$_2$ bilayer and trilayer structures created either  by random folding of flakes during exfoliation or by repeated deterministic stacking processes. We find that in all of our structures, the interlayer coupling is insufficient to modify the band structure or influence the crystal symmetry properties that govern valley physics in MoS$_2$, regardless of the relative stacking angle, which we determine by second-harmonic generation microscopy. However, a clear effect on the characteristic Raman modes is observable and allows us to qualitatively compare  the degree of interlayer coupling between different samples, and we find that it is significantly smaller in artificial structures created by deterministic stacking. Most likely, this is due to the long time interval in which the individual flakes are exposed to ambient conditions before the stacking process can be performed. During this time, adsorbates can accumulate at the exposed surfaces and act as a spacer layer in the artificial stack.  By contrast, the random folding of flakes occurs directly during the exfoliation process.
Even though there are no changes to the band structure, the photoluminescence emission in all of the artificial structures is strongly modified due to resonant energy transfer between the individual layers. At low temperatures, this mechanism is partially suppressed due to changes of the photocarrier dynamics. Our findings may play an important role in the design of future artificial two-dimensional crystal structures for opto-electronics: dipolar coupling between individual monolayers must be taken into account as a potential luminescence quenching mechanism, even if the layers are electronically decoupled.
\section*{Acknowledgements}
The authors  gratefully acknowledge financial support by the DFG via SFB689, KO3612/1-1 and GRK 1570.
\bibliography{MoS2}
\end{document}